\documentclass[doublecol]{epl2}
\usepackage{bm}
\usepackage{amsmath}

\title{Non-local Andreev reflection under ac bias}
\date{\today}

\author{D.S. Golubev\inst{1} \and A.D. Zaikin\inst{1,2} }
\shortauthor{D.S. Golubev \etal}
\institute{
  \inst{1} Forschungszentrum Karlsruhe, Institut f\"ur Nanotechnologie, 76021, Karlsruhe, Germany\\
\inst{2} I.E. Tamm Department of Theoretical Physics, P.N. Lebedev
Physics Institute, 119991 Moscow, Russia}


\abstract{We theoretically analyze non-local electron transport in
multi-terminal normal-metal-superconductor-normal-metal (NSN)
devices in the presence of an external ac voltage bias. Our
analysis reveals a number of interesting effects, such as, e.g.,
photon-assisted violation of balance between crossed Andreev
reflection (CAR) and elastic cotunneling (EC).  We demonstrate
that at sufficiently small (typically subgap) frequencies of an
external ac signal and at low temperatures the non-local
conductance of the NSN device turns negative implying that in this
regime CAR contribution to the non-local current dominates over
that of EC. Our predictions can be directly tested in future
experiments.}

\begin{document}

\maketitle


\section{Introduction}

The phenomenon of non-local (or crossed) Andreev reflection
\cite{BF,DF} is known to occur in multi-terminal hybrid
normal-metal-superconductor-normal-metal (NSN) proximity structures and
involves two subgap electrons entering a superconductor from two
different normal terminals and forming a Cooper pair there. Such
crossed Andreev reflection (CAR) manifests itself, e.g., in the
dependence of the current $I_1$ through the left NS interface of
an NSN structure on the voltage $V_2$ across the right NS
interface. As a result, the non-local conductance $G_{12}= - \partial
I_1/\partial V_2$ of an NSN device differs from zero and can be
detected experimentally. Various aspects of this intriguing
phenomenon have recently become a subject of intensive
investigations both in experiment
\cite{Beckmann,Teun,Venkat,Basel} and in theory, see, e.g.,
\cite{FFH,Martin,MF,BG,Morten,KZ06,GZ07,KZ07,LY,KZ08,Belzig,Koenig,Melin}
and further references therein.

It is important that CAR is not the only process which contributes
to the non-local conductance $G_{12}$. Another relevant process is
direct electron transfer between two normal terminals through the
superconductor. In the tunneling limit this process is just the
well known elastic cotunneling (EC). It was demonstrated
\cite{FFH} that in the lowest order in tunneling the contributions
from EC and CAR to $G_{12}$ exactly cancel each other in the limit
of low temperatures and voltages, i.e. the non-local conductance
$G_{12}$ tends to zero in this limit. In certain cases this observation
might complicate experimental identification of CAR
in NSN structures with weakly transmitting interfaces.

Despite such possible complications
in experiments \cite{Beckmann,Teun,Venkat,Basel} the
non-local signal $G_{12}$ was successfully detected both at high
and low temperatures showing a rich structure of non-trivial
features. Some of these features are currently not yet fully
understood and are still waiting for their adequate theoretical
interpretation. These observations -- along with
various theoretical predictions -- also demonstrate that the exact
cancellation between EC and CAR contributions \cite{FFH} can be
violated in a number of ways. One of them is simply to lift the spin
degeneracy in the problem, e.g., by considering NSN structures
with spin-active interfaces \cite{KZ07} or by using ferromagnets
(F) instead of normal electrodes \cite{MF}. Experiments with FSF
structures \cite{Beckmann} directly demonstrated the dependence of
the non-local conductance $G_{12}$ on the polarization of
F-electrodes.

In a spin degenerate case the non-local conductance $G_{12}$ does
not vanish beyond the tunneling limit \cite{KZ06,KZ07}, i.e. the
cancellation between EC and CAR terms is effectively eliminated
due to higher order electron tunneling processes which become
significant at higher barrier transmissions. This theory predicts
{\it positive} non-local conductance implying that direct electron
transfer should always dominate over CAR at higher barrier
transmissions. Furthermore, for ballistic NSN structures CAR was
predicted to vanish completely in the limit of fully open NS
interfaces \cite{KZ06,KZ07}.

It is worth pointing out that both positive and {\it negative}
non-local signals have been detected in multi-terminal NSN devices
\cite{Beckmann,Teun,Basel}. It was argued \cite{LY} that negative
non-local currents could possibly be attributed to the effect of
electron-electron interactions. In the presence of strong Coulomb
interaction negative non-local conductance
was obtained in single-level quantum dots coupled to
normal and superconducting electrodes \cite{Koenig}.
In general, this important issue deserves a
detailed theoretical investigation which should account for
non-trivial interplay between disorder and Coulomb interaction.
Previously a similar analysis was developed for (local) Andreev
reflection in NS hybrid structures \cite{Z,HHK,GZ06}. This
analysis revealed a number of interesting features which can also
significantly affect non-local properties of NSN devices.

Leaving this important issue for future investigations,
here we address a somewhat different situation. Namely, we will
study both EC and CAR processes in NSN structures in the presence
of an external ac bias. We will demonstrate that application of an
ac electromagnetic field to such structures lifts exact
cancellation between EC and CAR contributions already in the
lowest order in barrier transmissions. Under certain conditions
CAR can dominate over EC, in which case the non-local conductance
of the system $G_{12}$ turns negative.

\section{Non-local currents in NSN devices}

We will analyze the behavior of multi-terminal NSN structures. An
example of such a structure is schematically shown in Fig. 1: A
long superconducting wire with BCS order parameter $\Delta (T)$
(connected to bulk superconducting terminals) is coupled to two
normal metals via tunnel barriers with resistances $R_1$ and
$R_2$. We also assume that time dependent voltages $V_1$ and $V_2$
are applied to normal metal terminals.

The Hamiltonian of this system can be expressed in the form
\begin{eqnarray}
H=H_1+H_2+H_S+H_{1,T}+H_{2,T},
\end{eqnarray}
where
\begin{equation}
H_{r}=\sum_{\alpha=\uparrow,\downarrow}\sum_k \big(\epsilon_{k} -
eV_r(t)\big) c^\dagger_{r,k,\alpha}c_{r,k,\alpha},
\end{equation}
are the Hamiltonians of the normal leads ($r=1,2$),
\begin{equation}
H_S=\sum_k \big[\xi_k (a_{k,\uparrow}^\dagger a_{k,\uparrow} +
a_{k,\downarrow}^\dagger a_{k,\downarrow} ) +\Delta
(a^\dagger_{k,\uparrow}a^\dagger_{k,\downarrow} +
a_{k,\downarrow}a_{k,\uparrow}) \big]
\end{equation}
is the Hamiltonian of the superconducting electrode and
\begin{equation}
H_{r,T}=\sum_{\alpha=\uparrow,\downarrow}\sum_{k,p} \big[ t_{kp}
c_{r,k,\alpha}^\dagger a_{\alpha,p} + \,{\rm c.c.}\, \big]
\end{equation}
are the tunnel Hamiltonians describing electron transfer between
normal terminals and a superconductor. In what follows we will restrict our
analysis to the spin-degenerate case and will also ignore
inelastic effects.

\begin{figure}
\begin{center}
\includegraphics[width=8cm]{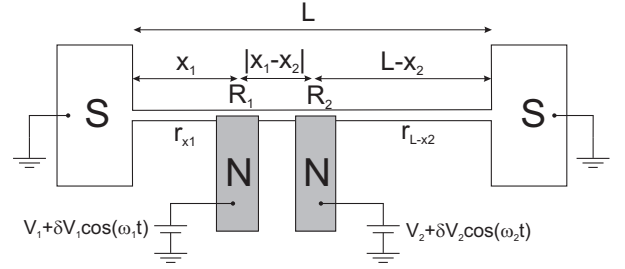}
\end{center}
\caption{Schematics of the system under consideration. A long
superconducting wire is coupled to two normal metal leads via the
two tunnel junctions with the resistances $R_1$ and $R_2$. }
\end{figure}

In order to evaluate the non-local current response to the applied
voltages we proceed within the standard perturbative approach in
the tunnel Hamiltonians $H_{1,T}$ and $H_{2,T}$ combined with the
Keldysh technique. The whole calculation is analogous to that
performed in Ref. \cite{FFH} with the only important difference
that here the applied voltages $V_{1,2}$ explicitly depend on
time. Calculating the current $I_1$ across the first barrier
perturbatively in the corresponding tunneling amplitudes, we obtain
\begin{eqnarray}
I_1(V_1,V_2) = I_1^{(0)}(V_1)  +I_{11}(V_1) +
I_{12}(V_2), \label{I1}
\end{eqnarray}
where the first two terms $I_1^{(0)}(V_1)$ and $I_{11}(V_1)$
represent the standard contributions to the current across NS
interface evaluated respectively in the first and second orders in
the interface transmission, while the term $I_{12}(V_2)$ describes
the non-local contribution to the current across the first
junction due to the presence of the voltage bias $V_2$ at the
second normal terminal. For these three contributions we find
\begin{eqnarray}
I_1^{(0)} &=& \frac{e}{2}\,{\rm tr}\,
\big[ -\check G_1\check t_1\check G_S\check t^\dagger_1\check\Lambda
+\check G_S\check t_1^\dagger\check G_1\check t_1\check\Lambda \big]_{t,t},
\label{I0}\\
I_{11}&=&
\frac{e}{2}\,{\rm tr}\,\big[ -\check G_1\check t_1\check G_S\check t_1^\dagger
\check G_1\check t_1\check G_S\check t^\dagger_1\check\Lambda
\nonumber\\ &&
+\,\check G_S\check t_1^\dagger G_1\check t_1\check G_S\check t_1^\dagger\check G_1\check t_1\check\Lambda \big]_{t,t},
\label{I11}\\
I_{12} &=& \frac{e}{2}\,{\rm tr}\,\big[ -\check G_1\check t_1\check G_S\check t_2\check G_2\check t_2^\dagger\check G_S\check t^\dagger_1\check\Lambda
\nonumber\\ &&
+\, \check G_S\check t_2\check G_2\check t_2^\dagger\check G_S\check t_1^\dagger\check G_1\check t_1\check\Lambda \big]_{t,t}.
\label{I12}
\end{eqnarray}
Eqs. (\ref{I1})-(\ref{I12}) express the current in terms of
unperturbed Green-Keldysh functions of normal and superconducting
terminals, respectively $\check G_{1,2}$, $\check G_S$. As
usually, these functions are $4\times 4$ matrices in Keldysh and
Nambu spaces. In equilibrium they depend only on the time
difference. For instance, for the Green-Keldysh function of the
superconducting terminal one has
$$\check G_S(t_1-t_2,x_1,x_2)=\int \frac{dE}{2\pi}\, e^{-iE(t_1-t_2)} \check G_S(E,x_1,x_2),$$
where
\begin{eqnarray}
\check  G_S(E,x_1,x_2)=\tilde G_{S}^R(E,x_1,x_2)\otimes \frac{\hat
1+\hat Q(E)}{2}\hat\sigma_z
\nonumber\\
+\, \tilde G_{S}^A(E,x_1,x_2)\otimes \frac{\hat 1-\hat
Q(E)}{2}\hat\sigma_z,
\end{eqnarray}
$\hat\sigma_z$ is the Pauli matrix and the $2\times 2$ matrices
$\tilde G_{S}^{R,A}(E,x_1,x_2)$ read
\begin{eqnarray}
 \tilde G_{S}^{R,A}(E,x_1,x_2) \equiv
\left(\begin{array}{cc} G_{S}^{R,A}(E) & F_{S}^{+,R,A}(E) \\
F_{S}^{R,A}(E) & G_{S}^{+,R,A}(E) \end{array}\right)=
\nonumber\\
\sum_n \left(
\begin{array}{cc}
\frac{(E+\xi_n\pm i0)\chi_n(x_1)\chi_n^*(x_2)}{(E\pm
i0)^2-\Delta^2-\xi_n^2} &
\frac{\Delta_j\chi_n(x_1)\chi_n(x_2)}{(E\pm i0)^2-\Delta^2-\xi_n^2} \\
\frac{\Delta_j\chi_n^*(x_1)\chi_n^*(x_2)}{(E\pm
i0)^2-\Delta^2-\xi_n^2} & \frac{(E-\xi_n\pm
i0)\chi_n(x_1)\chi_n^*(x_2)}{(E\pm i0)^2-\Delta^2-\xi_n^2}
\end{array}
\right). \nonumber
\end{eqnarray}
Here we have defined the normal state electron wave functions in the
superconducting  electrode $\chi_n(x)$, and
the matrix
\begin{eqnarray}
\hat Q(E)&=&\left(\begin{array}{cc} 1-2n(E) & 2n(E) \\
2-2n(E) & -1+2n(E)   \end{array}\right),
\end{eqnarray}
where $n(E)=1/(1+\exp(E/T))$ is the Fermi function. The matrix
$\hat Q(E)$ satisfies the normalization condition $\hat Q^2(E)=1$.
The Green-Keldysh functions for N-terminals $\check G_{1,2}$ are
defined analogously, one should just set $\Delta=0$.

Let us also note that the trace operation in Eqs.
(\ref{I0})-(\ref{I12}) includes taking the trace in $4\times 4$
Keldysh-Nambu space together with the convolution over intermediate
coordinates and times. The subscript $_{t,t}$ implies that the
outer times in the products of the Green functions should be equal
and no integration over these times should be performed.
Eqs. (\ref{I0})-(\ref{I12}) also involve the $4\times 4$
matrices $\check\Lambda$ and $\check t_{1,2}$. The matrix $\check\Lambda$
is diagonal with the following matrix elements:
$\Lambda_{11}=\Lambda_{44}=-1$, $\Lambda_{22}=\Lambda_{33}=1$.
The matrices $\check t_{1,2}$ read
\begin{eqnarray}
\check t_j=\left(
\begin{array}{cccc}
-t_je^{-i\varphi_j} & 0 & 0 & 0 \\
0 & t_j e^{-i\varphi_j} & 0 & 0 \\
0 & 0 & t_je^{i\varphi_j} & 0 \\
0 & 0 & 0 & -t_je^{i\varphi_j}
\end{array}\right),
\end{eqnarray}
where
$$\varphi_{1,2}(t)=\int^t dt' \, eV_{1,2}(t')$$
are the time dependent phases across the tunnel barriers.

The second order contributions to the current (\ref{I11}),
(\ref{I12}) describe several different physical effects. The
difference between these effects can be illustrated, e.g., in the
course of averaging of the Green functions over disorder. For
instance, the term (\ref{I11}) contains the product of four Green
functions $\check G_1\check G_S\check G_1\check G_S$. While
averaging, one can perform the perturbation theory in $1/g_S,$
$1/g_N$, where $g_{S,N}=2\pi / e^2r_{S,N}$ are dimensionless
conductances of superconducting and normal leads (see Fig. 1).
Keeping the leading and next to the leading contributions, one
finds
\begin{eqnarray}
\langle\check G_1\check G_S\check G_1\check G_S \rangle= A_0+A_1+A_2,
\end{eqnarray}
where
\begin{eqnarray}
A_0&=&\langle\check G_1\rangle\langle\check G_S\rangle\langle\check G_1\rangle\langle\check G_S \rangle
\end{eqnarray}
is the leading contribution $\propto (1/g_{S})^0,(1/g_N)^0$, and
\begin{eqnarray}
A_1&\propto &
\overline{\langle\check G_1 \check G_1\rangle}\langle\check G_S\rangle\langle\check G_S \rangle
\propto 1/g_N,
\nonumber\\
A_2 &\propto &
\langle\check G_1\rangle\langle\check G_1\rangle\overline{\langle\check G_S \check G_S\rangle}
\propto 1/g_S
\end{eqnarray}
are the first order corrections. Here the $\overline{\langle\check
G_{j} \check G_j\rangle} =\langle\check G_{j} \check
G_j\rangle-\langle\check G_{j}\rangle\langle \check G_j\rangle $
are irreducible averages which, in general, contain both diffusons
and Cooperons. { The average $A_0$ yields the standard subgap
current \cite{BTK} which can be disregarded in the limit of small
barrier transmissions considered here. The contribution $A_1$
describes Andreev reflection enhanced by disorder in the N-metal
\cite{VZK,HN,Z}. This contribution is also omitted here. For the
local current $I_{11}$ this approximation is justified ($i$) at
energies (e.g. $T$ or $\omega$) well above the Thouless energy of
the N-terminal or ($ii$) for sufficiently strong dephasing in the
N-metal or ($iii$) for $g_N\gg g_S$. Furthermore, this
approximation does not affect the lowest order in tunneling
contribution $\propto 1/R_1R_2$ to the non-local current $I_{12}$
at all. Thus, below we will keep only the terms $\propto A_2$ in
Eq. (\ref{I11}) and retain averages of the type $\langle\check
G_1\rangle \overline{\langle\check G_S\check G_S\rangle}\langle
G_2 \rangle$ in Eq. (\ref{I12}).}

Evaluating the traces in Eqs. (\ref{I0})-(\ref{I12}) and assuming
that exact particle-hole symmetry remains preserved, we get
\begin{eqnarray}
I_1^{(0)} &=& - \frac{1}{ e N_S R_1} \int dt' \,
G_{1}^R(t-t',x_1,x_1)  h(t-t')
\nonumber\\ &&\times\,
\sin[\varphi_1(t)-\varphi_1(t')],
\label{I0n}\\
I_{11}&=& \frac{i}{2e^3N_S^2R_1^2}\int dt_1dt_2
\nonumber\\ &&\times\,
\big[ G_{S}^R(t-t_1,x_1,x_1)G_{S}^A(t_2-t,x_1,x_1)
\nonumber\\ &&
-\, F_{S}^R(t-t_1,x_1,x_1)F_{S}^A(t_2-t,x_1,x_1) \big]
\nonumber\\ &&\times\,
n(t_1-t_2)\sin\big[\varphi_1(t_1)-\varphi_1(t_2)\big],
\label{I11n}\\
I_{12} &=& \frac{i}{2e^3N_S^2R_1R_2}\int dt_1dt_2
\nonumber\\ &&\times\,
\big[ G_{S}^R(t-t_1,x_1,x_2)G_{S}^A(t_2-t,x_2,x_1)
\nonumber\\ &&
-\, F_{S}^R(t-t_1,x_1,x_2)F_{S}^A(t_2-t,x_2,x_1) \big]
\nonumber\\ &&\times\,
n(t_1-t_2)\sin\big[\varphi_2(t_1)-\varphi_2(t_2)\big],
\label{I12n}
\end{eqnarray}
where we defined
$$h(t) = \int \frac{dE}{2\pi}\, e^{-iEt}\big(1-2n(E)\big).$$

Next we assume that all the electrodes in our system are
diffusive. In this case we need to average the above expressions
over disorder. This averaging is accomplished with the aid of the
following rules
\begin{eqnarray}
&& \langle G_{1}^R(E,x_1,x_1) \rangle = -\pi i N_S
\nonumber\\ &&\times\,
\left( \frac{|E|\theta(|E|-\Delta)}{\sqrt{E^2-\Delta^2}}
-i \frac{E\theta(\Delta-|E|)}{\sqrt{\Delta^2-E^2}} \right),
\end{eqnarray}
\begin{eqnarray}
&& \langle G_{S}^R(E_1,x_1,x_2)G_{S}^A(E_2,x_2,x_1)\rangle =
\nonumber\\ &&
 =\,\pi N_S\bigg( 1+\frac{E_1E_2}{W(E_1)W^*(E_2)} \bigg)
\nonumber\\ &&\times\,
{\cal D}_S\big( W(E_1)-W^*(E_2),x_1,x_2 \big),
\end{eqnarray}
\begin{eqnarray}
&& \langle F_{S}^R(E_1,x_1,x_2)F_{S}^A(E_2,x_2,x_1)\rangle =\pi N_S
\nonumber\\ &&\times\, \frac{\Delta^2}{W(E_1)W^*(E_2)} {\cal
D}_S\big( W(E_1)-W^*(E_2),x_1,x_2 \big).
\end{eqnarray}
Here $N_S$ is the normal density of states in the superconducting
electrode,
\begin{eqnarray}
W(E)&=&{\rm sign}\, E\,\theta(|E|-\Delta)\sqrt{E^2-\Delta^2}
\nonumber\\ &&
+\,i\theta(\Delta-|E|)\sqrt{\Delta^2-E^2},
\end{eqnarray}
and ${\cal D}_S(\omega,x.x')$ is the diffuson, which satisfies the
equation
\begin{eqnarray}
\left( -i\omega -D_S\nabla^2 \right){\cal D}_S(\omega,x,x')=\delta(x-x'),
\end{eqnarray}
where $D_S$ is the diffusion coefficient in the superconducting electrode.
Here we assume that the time reversal symmetry is maintained in
our problem, therefore we do not need to distinguish between the diffuson
and the Cooperon. Then we obtain
\begin{eqnarray}
 I_1^{(0)}  &=& - \frac{1}{ e N_S R_1} \int dt' \,
\big\langle G_{1}^R(t-t',x,x)\big\rangle  h(t-t')
\nonumber\\ &&\times\,
\sin[\varphi_1(t)-\varphi_1(t')],
\label{I0m}\\
 I_{11} &=& \frac{\pi i}{2e^3N_SR_1^2}\int dt_1dt_2\int \frac{dE_1dE_2dE_3}{(2\pi)^3}
\nonumber\\ &&\times\,
e^{-iE_1(t-t_1)-iE_2(t_2-t)-iE_3(t_1-t_2)}
\nonumber\\ &&\times\,
\left( 1+\frac{E_1E_2-\Delta^2}{W(E_1)W^*(E_2)} \right)\,n(E_3)
\nonumber\\ &&\times\,
{\cal D}_S\big( W(E_1)-W^*(E_2),x_1,x_1 \big)
\nonumber\\ &&\times\,
\sin\big[\varphi_1(t_1)-\varphi_1(t_2)\big],
\label{I110}\\
 I_{12} &=& \frac{\pi i}{2e^3N_SR_1R_2}\int dt_1dt_2\int \frac{dE_1dE_2dE_3}{(2\pi)^3}
\nonumber\\ &&\times\,
e^{-iE_1(t-t_1)-iE_2(t_2-t)-iE_3(t_1-t_2)}
\nonumber\\ &&\times\,
\left( 1+\frac{E_1E_2-\Delta^2}{W(E_1)W^*(E_2)} \right)\,n(E_3)
\nonumber\\ &&\times\,
{\cal D}_S\big( W(E_1)-W^*(E_2),x_1,x_2 \big)
\nonumber\\&&\times\,
\sin\big[\varphi_2(t_1)-\varphi_2(t_2)\big].
\label{I12m}
\end{eqnarray}
Together with Eq. (\ref{I1}) these general expressions define the
net current across the first barrier for arbitrary dependence of
the applied voltages $V_{1,2}(t)$ on time. Below we will analyze
these expressions in several specific limits.

\section{Linear ac response}

Let us choose the applied voltages in the form $V_j(t)=V_j+\delta
V_j e^{-i\omega_j t}$. Accordingly, the time dependent phases are
defined as
$$\varphi_j(t)=eV_jt + e\delta V_j\frac{1-e^{-i\omega_j
t}}{i\omega_j}.
$$
Substituting these expressions into the above results for the
current and expanding in the amplitudes of the ac signal $\delta
V_j$, we arrive at the correction to the current due to the
presence of ac bias:
\begin{eqnarray}
\delta I_1  =  G_{11}(\omega_1,V_1)\delta V_1 e^{-i\omega_1 t} - G_{12}(\omega_2,V_2)\delta V_2 e^{-i\omega_2 t}.
\end{eqnarray}
Here the choice of the minus sign in front of the second term is
just a matter of convention which ensures positive values of
$G_{12}$ at high bias voltages.

Let us first specify the nonlocal conductance $G_{12}(\omega,V_2)$.
It reads
\begin{eqnarray}
G_{12}= \int dE
\left( 1+\frac{E^2-\frac{\omega^2}{4}-\Delta^2}{W\left(E+\frac{\omega}{2}\right)W^*\left(E-\frac{\omega}{2}\right)} \right)
\nonumber\\ \times\,
\frac{{\cal D}_S\left(W\left(E+\frac{\omega}{2}\right)-W^*\left(E-\frac{\omega}{2}\right),x_1,x_2\right) }{ 8 e^2N_S R_1R_2 \omega}
\nonumber\\ \times\,
\bigg[ n\left(E-\frac{\omega}{2}+eV_2\right)+n\left(E-\frac{\omega}{2}-eV_2\right)
\nonumber\\
-\,n\left(E+\frac{\omega}{2}+eV_2\right)-n\left(E+\frac{\omega}{2}-eV_2\right)\bigg].
\label{G12}
\end{eqnarray}
The local ac conductance is given by the sum of two terms
\begin{eqnarray}
G_{11}(\omega,V)= G_{11}^{(0)}(\omega,V)  + \delta G_{11}(\omega,V),
\label{G11}
\end{eqnarray}
where
\begin{eqnarray}
G_{11}^{(0)} = \int \frac{dE }{  2R_1\omega}
\left[ \frac{|E|\theta(|E|-\Delta)}{\sqrt{E^2-\Delta^2}}
-i \frac{E\theta(\Delta-|E|)}{\sqrt{\Delta^2-E^2}} \right]
\nonumber\\   \times\,
\big[ n(E-\omega-eV_1) + n(E-\omega+eV_1)
\nonumber\\
-\, n(E-eV_1) - n(E+eV_1)    \big],
\label{G110}
\end{eqnarray}
and the correction $\delta G_{11}(\omega,V_1)$ is given by Eq.
(\ref{G12}) with a simple replacement  ${\cal
D}_S\left(\omega,x_1,x_2\right)$ $\to {\cal
D}_S\left(\omega,x_1,x_1\right)$.

Eqs. (\ref{G12})-(\ref{G11}) apply for any sample geometry. Let us
now specify these equations for the NSN structure illustrated in
Fig. 1. Provided the superconductor can be treated as a
sufficiently thin wire with cross-section $s$, the diffuson takes
the form
\begin{eqnarray}
{\cal D}_S(\omega,x,x') &=&
\frac{\cosh\left[\sqrt{\frac{-i\omega}{D_S}}(L-|x-x'|)\right]}
{2sD_S\sqrt{\frac{-i\omega}{D_S}}\sinh\left[\sqrt{\frac{-i\omega}{D_S}}\,L\right]}
\nonumber\\ &&
-\,\frac{\cosh\left[\sqrt{\frac{-i\omega}{D_S}}(L-x-x')\right]}
{2sD_S\sqrt{\frac{-i\omega}{D_S}}\sinh\left[\sqrt{\frac{-i\omega}{D_S}}\,L\right]}.
\end{eqnarray}
Substituting this expression into Eqs. (\ref{G12})-(\ref{G11}) we
arrive at our final results.

In the zero frequency limit $\omega \to 0$ the above general
expressions reduce to
\begin{eqnarray}
 G_{11}(0,V_1) = \frac{1}{eR_1}\int  \frac{|E|\theta(|E|-\Delta)}{\sqrt{E^2-\Delta^2}}
\frac{dE}{4T_1\cosh^2\frac{E-eV_1}{2T_1}}
\nonumber\\
-\,\frac{r_{x_1}r_{L-x_1}}{2r_LR_1^2}\bigg(2-\tanh\frac{\Delta-eV_1}{2T}
-\tanh\frac{\Delta+eV_1}{2T}\bigg),
\label{G11m}
\end{eqnarray}
\begin{eqnarray}
G_{12}(0,V_2) =
\frac{r_{x_1}r_{L-x_2}}{2r_LR_1R_2}\bigg(2-\tanh\frac{\Delta-eV_2}{2T}
\nonumber\\
-\,\tanh\frac{\Delta+eV_2}{2T}\bigg),
\label{G12m}
\end{eqnarray}
where the resistances $r_L$, $r_{x_1}$, $r_{L-x_1}$ and
$r_{L-x_2}$ are the normal state resistances of the corresponding
segments of the superconducting wire as it is illustrated in Fig.
1. We observe that at $T \to 0$ and $eV_2 < \Delta$ the non-local
conductance $G_{12}$ (\ref{G12m}) tends to zero in agreement with
\cite{FFH}.

\begin{figure}
\includegraphics[width=4cm]{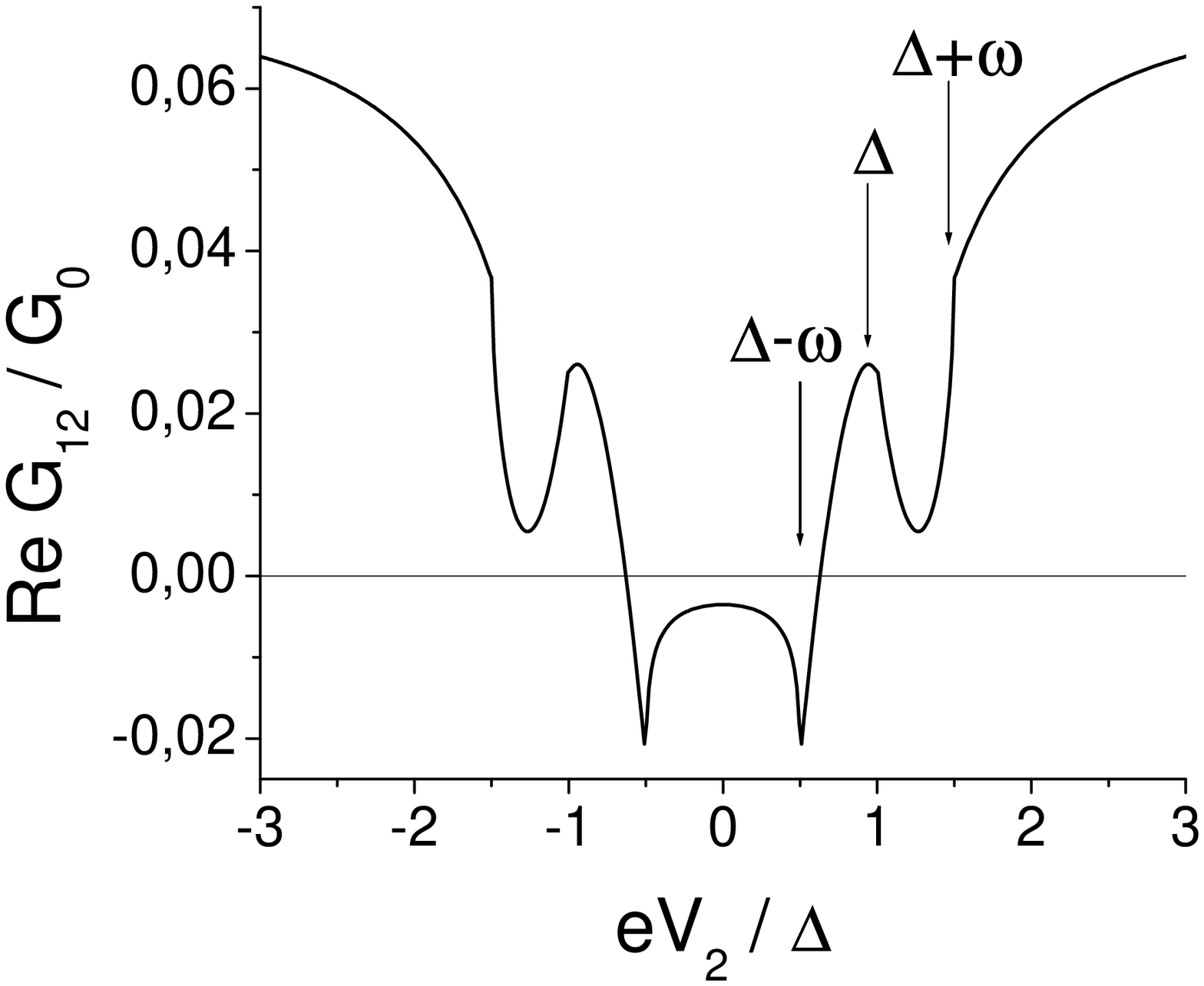}\hspace{0.5cm}
\includegraphics[width=4cm]{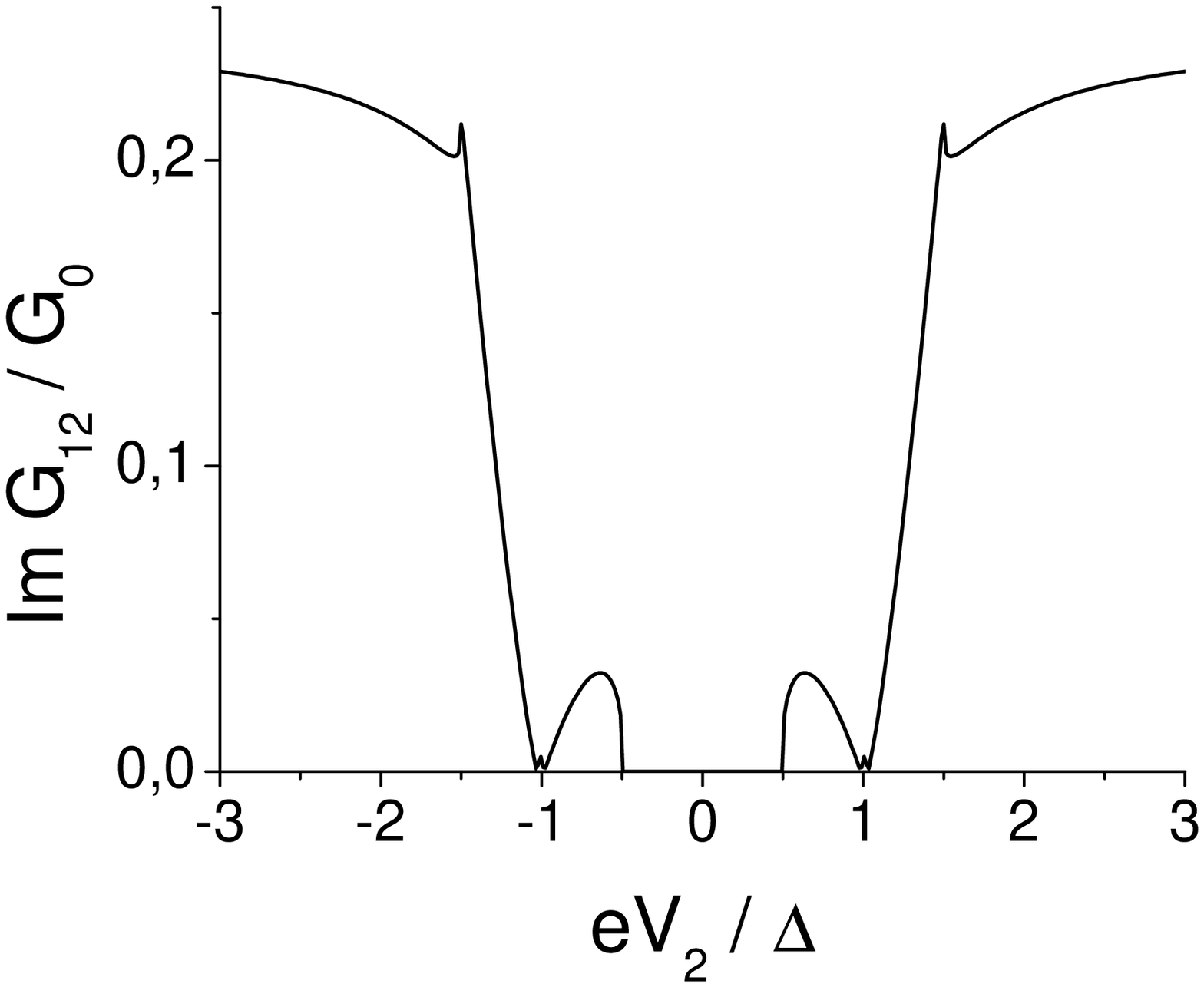}
\caption{Real and imaginary parts of the non-local conductance
$G_{12}(\omega,V_2)$ (respectively left and right panels) at zero
temperature and at $\omega=0.5\Delta$.  Here we fix $L=10\xi$,
$x_1=3\xi$, $x_2=4\xi$, where $\xi=\sqrt{D_S/\Delta}$ is the
superconducting coherence length. The the conductance values
are normalized by $G_0=r_{x_1}r_{L-x_2}/r_LR_1R_2$.}
\end{figure}

At non-zero frequencies the non-local conductance
$G_{12}(\omega,V_2)$ was evaluated numerically. The corresponding
results are presented in Figs. 2 and 3.
Fig. 2  illustrates the behavior of the non-local conductance
for a given value of $\omega=0.5\Delta$, $T \to 0$ and
different voltage bias values $V_2$. We
observe that an external ac bias lifts the balance between EC and
CAR processes essentially at all values $V_2$. On top of that,
at not very large bias voltages $V_2$ the real part of the
non-local conductance $G_{12}$ becomes {\it negative} implying
that in this regime
crossed Andreev reflection dominates over the elastic cotunneling.
At  $eV_2 = \pm \Delta \mp \omega$ this effect reaches its maximum
and $G_{12}$ shows clear dips related to the divergence of the superconducting
density of states at the gap energy. For  $|eV_2|> \Delta -\omega$ the
real part of the non-local conductance starts growing with increasing
bias and eventually
becomes positive at $|eV_2|> \Delta -\omega /2$. At $eV_2=\Delta$ the
non-local conductance Re$G_{12}$ has a peak which is again related to the
behavior of superconducting density of states in the vicinity of the gap
energy $\Delta$. Also the behavior of the imaginary part of the non-local
conductance Im$G_{12}$ (which differs from zero at  $|eV_2|> \Delta -\omega$)
turns out to be rather non-trivial, see Fig. 2.

\begin{figure}
\includegraphics[width=4cm]{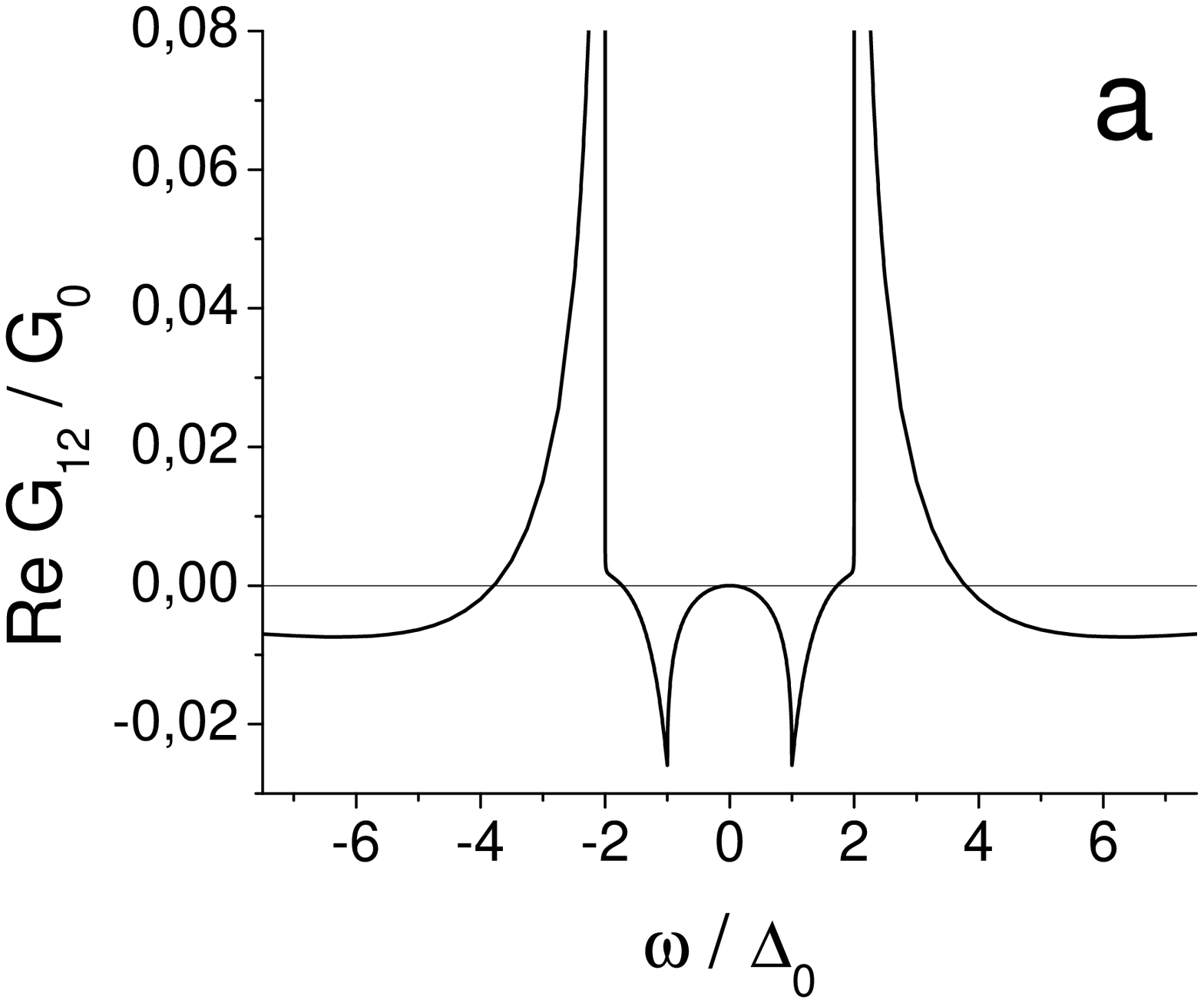}\hspace{0.5cm}
\includegraphics[width=4cm]{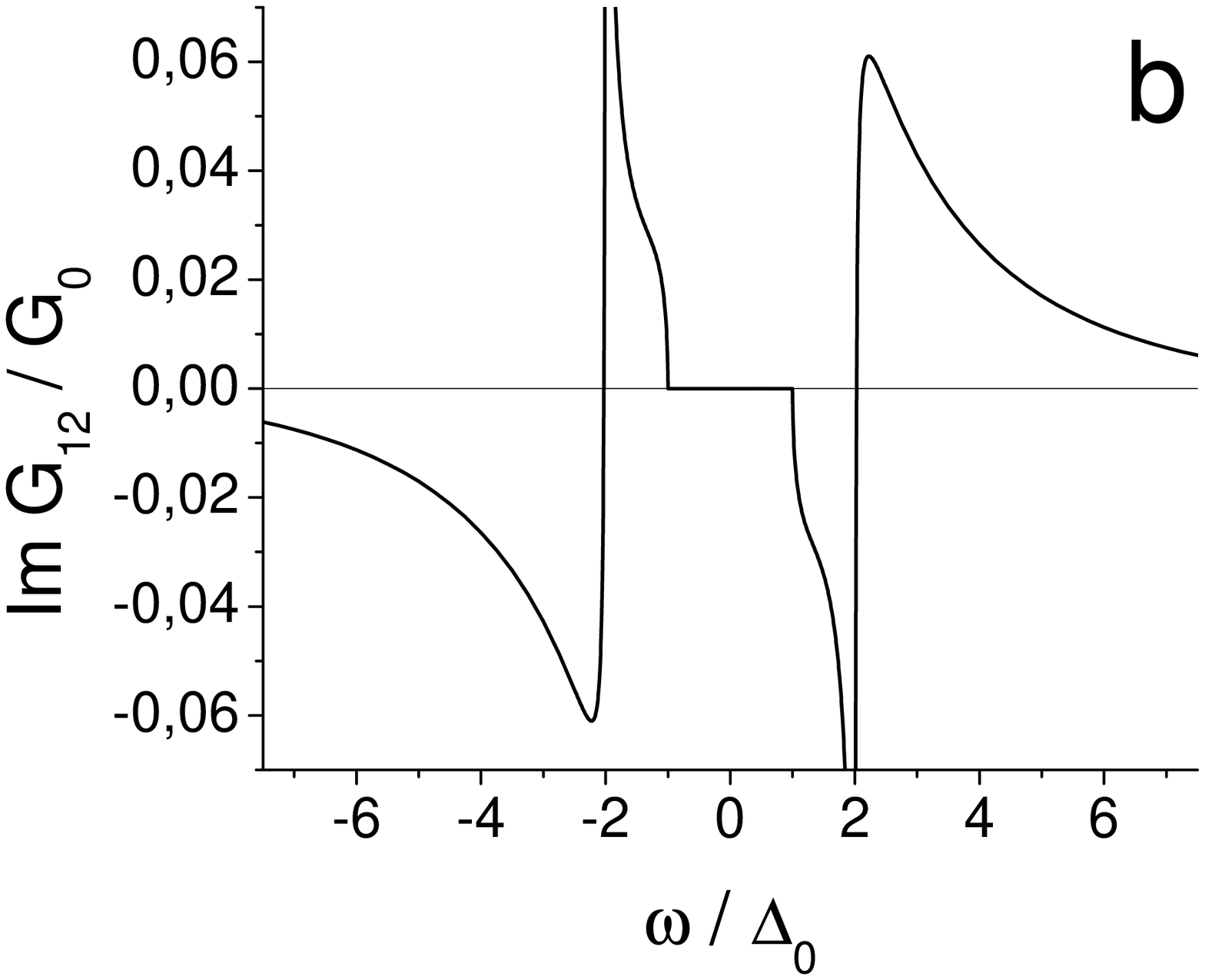} \\
\includegraphics[width=4cm]{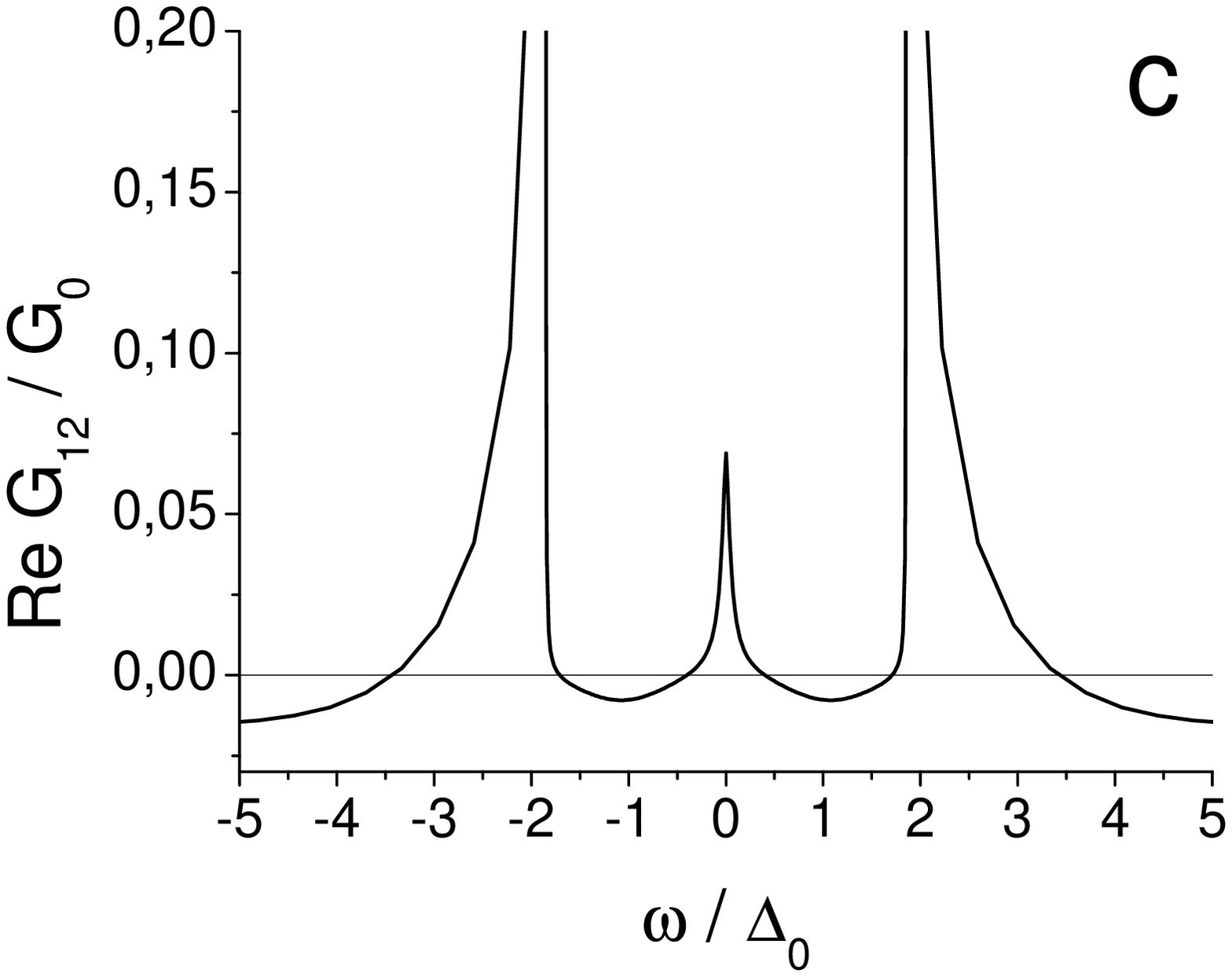}\hspace{0.5cm}
\includegraphics[width=4cm]{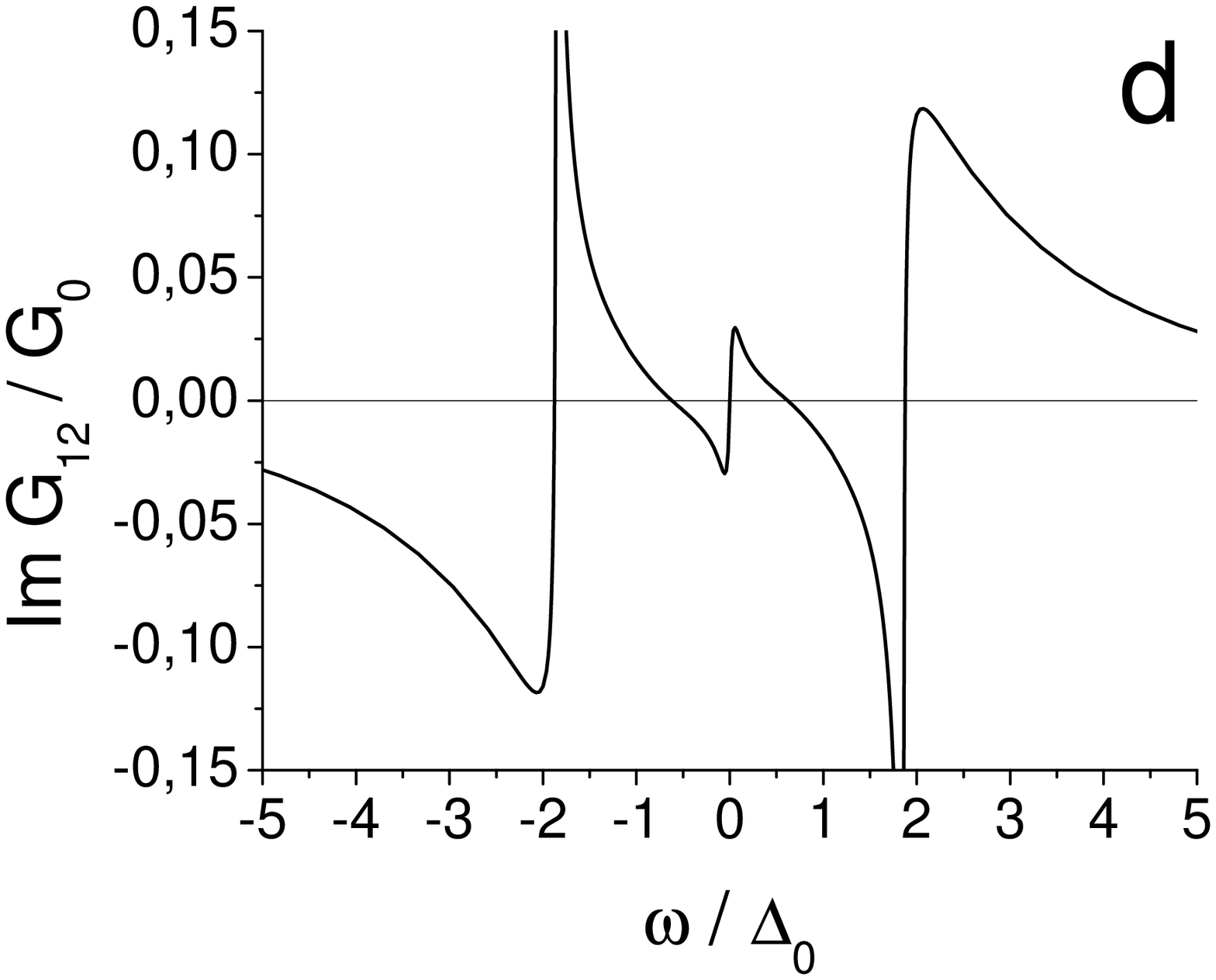}
\caption{The same as in Fig. 2 at zero bias voltage $V_2=0$ and
different ac bias frequencies $\omega$: (a) Re $G_{12}(\omega,0)$
at $T=0$, (b) Im $G_{12}(\omega,0)$ at $T=0$, (c) Re
$G_{12}(\omega,0)$ at $T=0.3\Delta_0$, where $\Delta_0$ is the value
of the superconducting gap at $T=0$, (d) Im $G_{12}(\omega,0)$
at $T=0.3\Delta_0$. Here we defined $\Delta_0 \equiv \Delta
(T=0)$.}
\end{figure}
\begin{figure}
\begin{center}
\includegraphics[width=5cm]{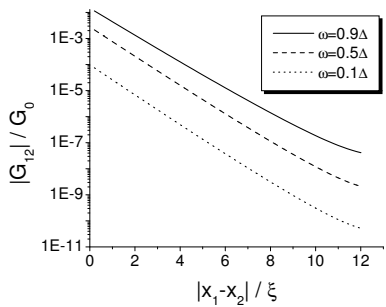}
\end{center}
\caption{{The absolute value of $G_{12}$ at $T=0$ as a function of
the distance between the barriers $|x_2-x_1|$. Similarly to the
case $\omega =0$ the dependence $|G_{12}|\sim
\exp(-|x_1-x_2|/\xi)$ is observed at subgap frequencies. Here we
set $V_2=0$, $L=20\xi$ and $x_1=7\xi$.}}
\end{figure}

The non-local conductance $G_{12}$ at zero dc bias $V_2=0$ and
different frequencies $\omega$ of an ac field is displayed in Fig.
3. We again observe that at $T=0$ and at frequencies $\omega$
below the superconducting gap $\Delta$ CAR dominates over EC thus
turning the real part part of the non-local conductance negative,
see Fig. 3a. This picture gets modified at non-zero $T$, in which
case EC contribution turns out to exceed that of CAR at
sufficiently small $\omega$, though at higher frequencies CAR can
still win over EC, cf. Fig. 3c. We also observe that a peak in the
real part of the non-local conductance develops at low frequencies
and non-zero $T$. The height of this peak { is essentially
determined by elastic cotunneling and} is defined by a simple
combination $G_{0}[1-\tanh\Delta/2T]$ (see Eq. (\ref{G12m})), 
while its width { is
roughly proportional to the Thouless energy $D_S/|x_1-x_2|^2$ and
slightly increases with temperature. At high enough temperatures
$T > \Delta(T)$ all singularities in the non-local conductance are
smeared and essentially disappear}.

{ At any subgap frequency $|\omega|<2\Delta$ the zero bias
non-local conductance decays exponentially with the distance
between the barriers. Similarly to the case $\omega=0$ for all
$|\omega|<\Delta$ we find $|G_{12}|\sim \exp(-|x_1-x_2|/\xi)$ as
it is illustrated in Fig. 4. }

\section{Non-linear response}

Finally let us turn to the non-linear non-local response. Choosing
the bias voltage in the form $V_2(t)=V_2+\delta V_2\cos\omega t$
we reconstruct the phase
$$\varphi_2(t)=eV_2t+\frac{e\delta
V_2}{\omega}\sin\omega t. $$ Substituting this function into Eq.
(\ref{I12m}) and averaging $\sin[\varphi_2(t_1)-\varphi_2(t_2)]$
over time, we obtain
\begin{eqnarray}
 \overline{\sin[\varphi_2(t_1)-\varphi_2(t_2)]}=
\sin[eV_2(t_1-t_2)]\bigg[J_0^2\left(\frac{e\delta V_2}{\omega}\right)
\nonumber\\
+2\sum_{n=1}^\infty J_n^2\left(\frac{e\delta
V_2}{\omega}\right)\cos[n\omega(t_1-t_2)]\bigg],
\end{eqnarray}
where $J_n(x)$ are the Bessel functions. Then the non-local
differential conductance takes the form
\begin{eqnarray}
-\frac{\partial I_{12}}{\partial V_2} =
J_0^2\left(\frac{e\delta V_2}{\omega}\right)G_{12}(0,V_2)
+\sum_{n=1}^\infty J_n^2\left(\frac{e\delta V_2}{\omega}\right)
\nonumber\\ \times\,
\left[G_{12}\left(0,V_2+\frac{n\omega}{e}\right)+G_{12}\left(0,V_2-\frac{n\omega}{e}\right)\right].
\label{G12non}
\end{eqnarray}
Here $G_{12}(0,V_2)$ is non-local conductance defined in Eq.
(\ref{G12m}). We observe that the non-local differential
conductance (\ref{G12non}) does not change its sign and
for non-zero $\omega$ remains
positive at all values of the bias voltage $V_2$.

\section{Concluding remarks}

Our analysis of the effect of an external ac bias on non-local
electron transport in multi-terminal NSN structures revealed a
number of interesting features which can be tested in future
experiments. It turns out that in the presence of an ac field
exact cancellation between zero-temperature EC and CAR
contributions to the non-local current is lifted already in the
lowest order in the transmissions of NS interfaces. { In the
presence of ac bias -- unlike in its absence -- the conductance
$G_{12}$ (\ref{G12}) is not anymore proportional to the density of
states in the S-metal (which vanishes at subgap energies thus
explaining the cancellation \cite{FFH}). Hence, no cancellation
between EC and CAR contributions to $G_{12}(\omega)$ should be
expected.} Considering these two contributions separately, we
observe that both EC and CAR processes get enhanced by external
radiation, however the subtle balance between them is eliminated.
As a result, at sufficiently small (typically subgap) frequencies
$\omega$ of the external ac signal the non-local conductance of
the NSN device $G_{12}$ turns negative implying that CAR
contribution wins over that of EC. { Although our present
analysis was carried out for diffusive structures our key
conclusions should also apply in the (quasi-)ballistic limit.}

The behavior described above can be realized in multi-terminal NSN
structures either under direct application of an ac bias or if the
geometry is such that an ac Josephson effect can occur somewhere
in the system in the presence of a dc voltage bias. This can be
the case, e.g. if a weak link (such as a pinhole or constriction)
is occasionally formed between different superconducting
terminals. In this case ac Josephson generation may occur under dc
bias effectively playing a role of an ac signal analyzed in our
work.

We also note that our results -- though indirectly -- could also
be of some relevance to the effect of electron-electron
interactions on the non-local properties of NSN devices. Indeed,
it is well known that this effect can generally be described by
(quantum) electromagnetic fields which mediate electron-electron
interactions. In this respect photon-assisted violation of balance
between EC and CAR (in favor of the latter process) considered
here might give a clue in which way Coulomb interaction could be
responsible for the negative non-local conductance observed in
recent experiments. However, a more elaborate calculation is
necessary to properly account for an interplay between disorder
and electron-electron interactions in multi-terminal NSN
structures. This calculation will be published elsewhere.

\end{document}